\documentclass[aps,pra,groupedaddress,twocolumn]{revtex4}
\usepackage{graphicx}
\usepackage{epsfig}
\usepackage{hyperref}
\usepackage{amssymb, amsmath, amsthm}

\newcommand{\norm}[1]{\left\Vert#1\right\Vert}
\newcommand{\abs}[1]{\left\vert#1\right\vert}
\newcommand{\set}[1]{\left\{#1\right\}}

\newcommand{\trace}[1]{\mbox{tr}\left( #1 \right)}

\newcommand{\refe}[1]{(\ref{#1})}

\newcommand{\UNICAMP}{Departamento de Matem\'atica Aplicada, Universidade Estadual de Campinas, Campinas, SP 13083-859,
Brazil}

\newcommand{\UFMG}{Departamento de F\'isica, ICEx, Universidade Federal de Minas Gerais, Av. Ant\^onio Carlos 6627, Belo Horizonte, MG 31270-901, Brazil}

\begin{document}

\title{Quantum State Tomography with incomplete data:\\Maximum Entropy and Variational Quantum Tomography\thanks{Partially supported by INCT-IQ, and FAPESP (Grants 2009/08027-0), FAPEMIG.}}

\author{D. S. Gon\c{c}alves}
\email[]{dgoncalves@ime.unicamp.br}
\affiliation{\UNICAMP}

\author{C. Lavor}
\affiliation{\UNICAMP}

\author{M. A. Gomes-Ruggiero}
\affiliation{\UNICAMP}

\author{A. T. Ces\'ario}
\affiliation{\UFMG}

\author{R. O. Vianna}
\affiliation{\UFMG}

\author{T. O. Maciel}
\email[]{thiago@fisica.ufmg.br}
\affiliation{\UFMG}
     
\date{\today}

\begin{abstract}
Whenever we do not have an informationally complete set of measurements, the estimate of a quantum state can not be uniquely determined. In this case, among the density matrices compatible with the available data, it is commonly preferred that one which is the most uncommitted with the missing information. This is the purpose of the Maximum Entropy estimation (MaxEnt) and the Variational Quantum Tomography (VQT). Here, we propose a variant of Variational Quantum Tomography and show its  relationship with Maximum Entropy methods in quantum tomographies with incomplete set of measurements. We prove their equivalence in case of eigenbasis measurements, and through numerical simulations we stress their similar behavior. Hence, in the modified VQT formulation we have an estimate of a quantum state as unbiased as in MaxEnt and with the benefit that VQT can be more efficiently solved by means of linear semidefinite programs. \\
\ \\
\noindent PACS number(s): 03.65.Wj, 03.67.-a
\end{abstract}


\maketitle


\section{Introduction}\label{introsec}
The task of estimating density matrices based on measurement results, or simply Quantum State Tomography (QST), is essential in quantum computation operations \cite{ncbook, clavor2011, maciel2012}. Besides being a subject valuable on its own,  QST is also used in quantum process tomography and in validation of quantum gates. However, it is known that the number of required measurements grows exponentially in the number of qubits. In this case, to take an informationally complete set of measurements becomes impractical in real experiments. Thus, it is important to apply methods that can deal with incomplete information. 

Whenever we have incomplete measurements, the state may not be uniquely determined by the available data, and suitable estimation methods which will not bias the undetected components are needed for correct interpretation and quantum diagnostics.
The Maximum Entropy estimation (MaxEnt) \cite{buzek1999} is usually adopted for this purpose. However this is not an easy task due to the nonlinearities in the reconstruction algorithm. Another possible approach, which avoids nonlinear difficulties, is the Variational Quantum Tomography (VQT) \cite{maciel2011}. VQT is cast as a linear semidefinite programming problem (SDP) \cite{klerkbook2002,boyd2004} and it searches for an estimate that is compatible with available data and minimizes a linear cost function: the sum of expected values of missing projectors. The similarities and differences between VQT and MaxEnt for quantum tomography with incomplete data were not analyzed before. 

Here, first we address the behavior of these two methods in how they set the probabilities related to missing measurements. Then, we propose a {\it modified} Variational Quantum Tomography and show that its behavior is quite similar to the Maximum Entropy approach, in the sense that it sets the unmeasured probabilities as uniform as possible. The advantage of the modified VQT is that it still can be formulated as  a linear semidefinite programming problem, which can be more efficiently solved than  non-linear optimization problems that arise from the MaxEnt formulation. 

This paper is organized as follows. Sec. \ref{vqtsec} briefly reviews the Variational Quantum Tomography and  Sec. \ref{maxentsec} covers the Maximum Entropy estimation.  In Sec. \ref{vqteigb}, we study how VQT and MaxEnt assign the unmeasured probabilities in the particular case of eigenbasis measurements. Then, we propose a change in the VQT formulation and show the relationship between this approach and the MaxEnt one in the case of incomplete but noise-free measurements. Sec. \ref{mlmesec} extends the application of MaxEnt to real experimental conditions, with the Maximum Likelihood assisted by Maximum Entropy (MaxLik-MaxEnt) approach \cite{rehacek2005}. In Sec. \ref{simsec}, through simulated incomplete experimental data, the behavior of MaxLik-MaxEnt and modified VQT approaches is illustrated. Firstly, we present noiseless data in order to stress the theoretical properties, and then we introduce two kinds of noise to assess the practical performance of the methods. Final remarks are done in Sec. \ref{finalsec}.

\section{Variational Quantum Tomography}\label{vqtsec}
In \cite{maciel2011}, Maciel et al. introduced the Variational Quantum Tomography to deal with incomplete set of measurements. The VQT formulation reads
\begin{equation}
\begin{aligned}
\underset{{\rho, \ \Delta}}{\text{minimize}} \ \ \ & \sum_{i \in {\cal I}} \Delta_i + \sum_{i \notin {\cal I}} \trace{E_i \rho} \\
\text{subject to} \ \ \ & \abs{\trace{E_i \rho} - f_i} \le \Delta_i f_i & \ \ i \in {\cal I}  \\
& \Delta_i \geq 0, & \ \\
& \trace{ \rho} \ = 1, & \  \\
& \rho \ \succeq 0,  \\
\end{aligned}
\label{vqt}
\end{equation}
where $\set{E_i}$ is the POVM set, $\set{f_i}$  the measured data, $\set{\Delta_i}$ the tolerances and ${\cal I}$ stands for 
the set of indexes of  measured data. The last two constraints guarantee that the estimated density matrix is
normalized and positive semidefinite.
 The other constraints are easy to understand. In the case of an ideal experiment (without noise), we require that the 
estimate for $\rho$ be compatible with the measured data, that is,
\begin{equation}
\label{linear_eq}
\trace{E_i \rho} = f_i, \ \ \forall i \in {\cal I}.
\end{equation}
However, in practice, the noise disturbs the measured expectation values so we allow that the above equations be
 violated by a small positive relative tolerance, such that
\begin{equation}
\label{linear_ineq}
\abs{\trace{E_i \rho} - f_i} \le \Delta_i f_i \ , \ \ \forall i \in {\cal I}.
\end{equation}
Of course we would like these tolerances to be minimal, which explains the first term in the objective function.
Note that the better the measurements, the smaller the $\Delta_i$, and for {\em ideal} noiseless measurements the
$\Delta_i$ are null. However, 
 in the incomplete data scenario, there may exist more than one estimate that minimizes those tolerances and satisfies the
 other constraints. Among all the acceptable solutions,  
VQT chooses the one which minimizes the expectation values of the 
unmeasured observables, namely:
\begin{equation}
\label{energy}
E =  \trace{ \left( \sum_{i \notin {\cal I}} E_i \right) \rho} = \sum_{i \notin {\cal I}} \trace{E_i \rho}.
\end{equation}
Therefore, $E$ is a {\em cost function} on the missing data.

One of the main virtues of VQT is that the formulation \refe{vqt} is a linear SDP 
(semidefinite programming problem \cite{klerkbook2002,boyd2004}) for which there are a lot of 
efficient methods \cite{helm1996,nemtodd2008} and some of them have polynomial computational 
complexity \cite{nemtodd2008}, which is desirable for scalability.

Another practical virtue, besides considering incomplete data, is that we can define an 
upper bound for  the variables $\Delta_i$. This allows us to control the quality of the fit and,
 moreover, it allows us to identify incompatible data through the infeasibility of the
 problem \refe{vqt}.

\section{Maximum Entropy estimation}\label{maxentsec}
The Maximum Entropy approach was introduced by Buzek et al. \cite{buzek1999}, 
in the context of quantum state tomography with incomplete data. 
The idea is to take as an estimate for $\rho$ the density matrix that maximizes the 
von Neumann entropy,
\begin{equation}
S(\rho) = -\trace{\rho \ln \rho},
\end{equation}
and is compatible with the available observed data. This estimate is obtained by solving the 
following optimization problem: 
\begin{equation}
\label{maxent}
\begin{aligned}
\underset{{\rho}}{\text{maximize}} \ \ \ & - \trace{\rho \ln \rho} \\
\text{subject to} \ \ \ & \trace{E_i \rho} \ = f_i, & \  i \in {\cal I} \ \\
    \ \ \ & \trace{\rho} \ = 1, & \  \\
& (\rho \ \succeq 0). & 
\end{aligned}
\end{equation}
Although we have to deal again with constraints in the semidefinite positive matrix space, it is possible to devise an explicit solution for \refe{maxent}, using the first order optimality conditions \cite{boyd2004, nocedal}. 

Applying the first order optimality conditions, we obtain
\begin{eqnarray}
\ln \rho + I + \lambda_0 I + \sum_{i \in {\cal I}} \lambda_i E_i & = & S, \nonumber \\
\trace{E_i \rho} & = & f_i, \  i \in {\cal I}  \\ 
\trace{\rho} & = & 1, \nonumber \\
\rho \succeq 0, & \ & S \succeq 0 \nonumber \\
\rho\,S & = & 0. \nonumber
\end{eqnarray}
Assuming that $\rho \succ 0$, we get
\begin{eqnarray}
\ln \rho + I + \lambda_0 I + \sum_{i \in {\cal I}} \lambda_i E_i & = & 0, \nonumber \\
\trace{E_i \rho} & = & f_i, \  i \in {\cal I} \label{kktmaxent} \\ 
\trace{\rho} & = & 1, \nonumber 
\end{eqnarray}
where $\lambda_i$ are the Lagrange multipliers associated to the equality constraints. From the first equation in  \refe{kktmaxent}, we obtain
\begin{equation}
\rho = \exp \left(  - \sum_{i \in {\cal I}} \lambda_i E_i - \lambda_0 I -I \right) \succ 0,
\end{equation}
and defining $1/\exp(-\lambda_0-1) = {\cal N} = \trace{ \exp \ \sum_{i} -\lambda_i E_i }$, we have
\begin{equation}
\label{mee}
\rho_{ME} = \frac{1}{{\cal N}} \exp \sum_{i \in {\cal I}} -\lambda_i E_i .
\end{equation}
Clearly, $\rho_{ME} \succ 0$ and $\trace{\rho_{ME}}=1$ due to the normalization constant ${\cal N}$. The Lagrange multipliers can be determined by solving the non-linear system of equations
\begin{equation}
\trace{E_j \rho_{ME}} \ = f_j, \ \  \ j \in {\cal I},
\end{equation}
that is,
\begin{equation}
\label{eqmaxent}
\trace{E_j \exp \sum_{i \in {\cal I}} -\lambda_i E_i} = {\cal N} f_j, \ \ \ j \in {\cal I}.
\end{equation}
In general, since we have noisy data $\set{f_i}$, we solve the following non-linear least-squares problem
\begin{equation}\label{nlsq}
\underset{\lambda}{\text{minimize}} \ \ \sum_{j \in {\cal I}} \left[\trace{E_j \exp \sum_{i \in {\cal I}} -\lambda_i E_i} - {\cal N} f_j \right]^2,
\end{equation}
instead of the non-linear equations \refe{eqmaxent}.

\section{VQT, Eigenbasis Measurements and Maximum Entropy}\label{vqteigb}
Since we are considering quantum tomography with incomplete measurements, an important point is
 how the methods assign the probabilities associated to unmeasured POVM elements.
In order to simplify our analysis, we will consider first the case of eigenbasis measurements in an ideal experiment (free of noise). Suppose we know that the true state $\rho$ can be written as (spectral decomposition):
\begin{equation}
\rho = \sum_{i \in {\cal I}} c_i P_i + \sum_{i \notin {\cal I}} c_i P_i,
\end{equation}
where $P_i$'s are $d$ orthonormal projectors onto the eigenspace of $\rho$. The task of tomography now is to determine the coefficients $c_i$'s (eigenvalues) based on observed data. Assuming that we have measured $m<d$ projectors ($d$ is the dimension of the Hilbert space), and that $\set{P_i}$ is an orthonormal set, it is easy to show that
\begin{equation}
\label{eigveq1}
c_i = \trace{P_i \rho}, \ \ \forall i,
\end{equation}
and since we have an ideal experiment, we also obtain
\begin{equation}
\label{eigveq2}
\trace{P_i \rho} = f_i, \ \ \forall i \in {\cal I},
\end{equation}
where $f_i$'s are the measured data. Due to the normalization constraint, we have that
\begin{equation}
\label{normeq}
\sum_{i \notin {\cal I}} c_i = 1 - \sum_{i \in {\cal I}} c_i = 1 - \sum_{i \in {\cal I}} f_i,
\end{equation}
and the constraint $\rho \succeq 0$ implies that $c_i \ge 0, \forall i$. 

Now let us consider the MaxEnt solution, given by \refe{mee}:
\begin{eqnarray}
\rho_{ME} & = &  \frac{1}{{\cal N}} \exp \left( \sum_{i \in {\cal I}} -\lambda_i P_i - \sum_{i \notin {\cal I}} 0 P_i \right)  \nonumber \\ 
          & = & \sum_{i \in {\cal I}} \frac{e^{-\lambda_i}}{{\cal N}} P_i + \frac{1}{{\cal N}} \sum_{i \notin I} P_i,
\end{eqnarray}
where $\lambda_i$ are the Lagrange multipliers related to the constraints \refe{eigveq2}. Since $\rho_{ME}$ must satisfy those constraints, we have
\begin{equation}
f_j =  \trace{P_j \left( \sum_{i \in {\cal I}} \frac{e^{-\lambda_i}}{{\cal N}} P_i + \frac{1}{{\cal N}} \sum_{i \notin I} P_i \right) } = \frac{e^{-\lambda_j}}{{\cal N}}.
\end{equation}
Thus, $e^{-\lambda_i}/{\cal N} = f_i = c_i, \ \ \forall i \in {\cal I}$, as we expected. Moreover, as $\trace{\rho_{ME}} = 1$, we obtain
\begin{equation}
\sum_{i \in {\cal I}} f_i + \frac{d-m}{{\cal N}} = 1,
\end{equation}
which implies that, for the unmeasured coefficients,
\begin{equation}
\label{uvec}
c_i = \frac{1}{{\cal N}} = \frac{1 - \sum_{j \in {\cal I}} f_j}{d - m}, \ \  \ \forall i \notin {\cal I}.
\end{equation}
In other words, the MaxEnt solution
\begin{equation}
\label{meeb}
\rho_{ME} = \sum_{i \in {\cal I}} f_i P_i + \sum_{i \notin {\cal I}} \left( \frac{1 - \sum_{j \in {\cal I}} f_j}{d - m} \right) P_i,
\end{equation}
uniformly distributes the remainder $1 - \sum_{j \in {\cal I}} f_j$ among the other coefficients $c_i, \forall i \notin {\cal I}$.  

Now let us compare this solution with the VQT solution. Considering the formulation \refe{vqt}, and applying  equations \refe{eigveq1},\refe{eigveq2},\refe{normeq}, we obtain
\begin{equation}
\begin{aligned}
\underset{c_i, \forall i \notin {\cal I}}{\text{minimize}} \ \ \ & \sum_{i \notin {\cal I}} c_i \\
\text{subject to} \ \ \ & \sum_{i \notin {\cal I}} c_i = 1 - \sum_{i \in {\cal I}} f_i,  \  \\
\ & c_i \ge 0 , \ \forall i \notin {\cal I}. & \ \\
\end{aligned}
\label{vqteb}
\end{equation}
Since we are assuming an ideal experiment, $c_i = \trace{P_i \rho} = f_i, \ \forall i \in {\cal I}$, we have $\Delta_i = 0, \forall i \in {\cal I}$. Furthermore, the variables of the problem \refe{vqteb} are $c_i, \forall i \notin {\cal I}$, and any feasible solution of \refe{vqteb} is also optimal, because the objective function is the same as the left hand side of the first constraint.
Thus we can conclude that 
the solution of VQT is expressed as
\begin{equation}
\rho_{VQT} = \sum_{i \in {\cal I}} f_i P_i + \sum_{i \notin {\cal I}} c_i P_i.
\end{equation}
Therefore, there is no constraint or penalty in the objective function that forces $c_i, i \notin {\cal I}$, agree with those of MaxEnt solution.

In order to guarantee that the VQT solution agrees with the MaxEnt solution, at least in the ideal eigenbasis case, we propose a change in the VQT formulation \refe{vqt}. Let us define the vector $\tilde{c}$, of size $d^2-m$, with components
\begin{equation}
\tilde{c}_i = \trace{E_i \rho}, \ \ \forall i \notin {\cal I}.
\end{equation}
Considering that $E_i$ are usually POVM elements (or projectors), we will assume that $\tilde{c}_i \ge 0$. Thus we have that
\begin{equation}
\norm{\tilde{c}}_1 = \sum_{i \notin {\cal I}} \abs{\trace{E_i \rho}} = \sum_{i \notin {\cal I}} \trace{E_i \rho} = \trace{H\rho} = E.
\end{equation}
Our proposal consists in using
\begin{equation}
\norm{\tilde{c}}_{\infty} = \max_{i \notin {\cal I}} \abs{\trace{E_i \rho}} = \max_{i \notin {\cal I}} \ \trace{E_i \rho},
\end{equation}
instead of $\norm{\tilde{c}}_1$ in the objective function of the problem \refe{vqt}. Whenever the sum of the components of $\tilde{c}$ is fixed, minimizing $\norm{\tilde{c}}_{\infty}$ promotes a more uniform distribution of these coefficients. 

In this case, the VQT$_{\infty}$ (VQT with $\norm{.}_{\infty}$) formulation becomes
\begin{equation}
\begin{aligned}
\underset{\rho, \ \Delta}{\text{minimize}} \ \ \ & \sum_{i \in {\cal I}} \Delta_i + \max_{i \notin {\cal I}} \ \trace{E_i \rho} \\ 
\text{subject to} \ \ \ & \abs{\trace{E_i \rho} - f_i} \le \Delta_i f_i & \ \ i \in {\cal I}  \\
& \Delta_i \ \geq 0, & \  \\
& \trace{ \rho} \ = 1, & \  \\
& \rho \ \succeq 0, & \ \\
\end{aligned}
\label{vqtinf}
\end{equation}
and for the ideal eigenbasis case, the equivalent of \refe{vqteb} is
\begin{equation}
\begin{aligned}
\underset{c_i, \forall i \notin {\cal I}}{\text{minimize}} \ \ \ & \max_{i \notin {\cal I}} c_i \\
\text{subject to} \ \ \ & \sum_{i \notin {\cal I}} c_i = 1 - \sum_{i \in {\cal I}} f_i,  \  \\
\ & c_i \ge 0 , \ \forall i \notin {\cal I}. & \ \\
\end{aligned}
\label{vqtebinf}
\end{equation}
Notice that the problem \refe{vqtebinf} has a unique solution that corresponds to the uniform distribution of the remainder $1 - \sum_{i \in {\cal I}} f_i$ among the coefficients $c_i, \forall i \notin {\cal I}$, which  coincides with the MaxEnt solution in this case.  

Despite the agreement with MaxEnt in the eigenbasis case,
 the proposed modification does not turn the problem \refe{vqtinf} harder than \refe{vqt}. 
This is true because minimizing $\norm{\tilde{c}}_{\infty}$ is equivalent to minimize some 
auxiliary variable $\delta$ subject to $\abs{\tilde{c}_i} \le \delta, \ \forall i \notin {\cal I}$.
Therefore, we have again a linear SDP:
\begin{equation}
\begin{aligned}
\underset{\rho, \ \Delta,\ \delta}{\text{minimize}} \ \ \ & \sum_{i \in {\cal I}} \Delta_i + \delta \\ 
\text{subject to} \ \ \ & \abs{\trace{E_i \rho} - f_i} \le \Delta_i f_i & \ \ i \in {\cal I}  \\
\  \ \ \ & \trace{E_i \rho} \le \delta & \ \ i \notin {\cal I}  \\
& \Delta_i \ \geq 0, & \  \\
& \trace{ \rho} \ = 1, & \  \\
& \rho \ \succeq 0. & \ \\
\end{aligned}
\label{vqtinf0}
\end{equation}
Although the equivalence between the solution of VQT formulation \refe{vqtinf0} and the one of
 MaxEnt is true only for the eigenbasis case, the relationship between these two problems is clear:
 both of them, each in its  own way, try to set the unmeasured probabilities the most uniformly
 as possible.

\section{Noisy data and the MaxLik-MaxEnt approach}\label{mlmesec}
To handle noisy data in the MaxEnt approach, we review a statistically based method that is 
employed in quantum tomography \cite{hradil2004, rehacek2007,james2001}: the Maximum Likelihood estimation (MaxLik) \cite{lc1998}. MaxLik searches an estimation for a density matrix such that the observed data are most likely:
\begin{equation}
\label{mle}
\begin{aligned}
\underset{\rho}{\text{maximize}} \ \ \ & {\cal L}(\rho | n) \equiv f(n | \rho) \\
\text{subject to} \ \ \ & \trace{\rho} \ = 1, & \  \\
& \rho \ \succeq 0, & 
\end{aligned}
\end{equation}
where $f(n | \rho)$ is the joint probability density function of the observed data $n$ given $\rho$. 

Of course the objective function of \refe{mle} depends on the parametric model assumed. One can use, for example, a multinomial distribution, as considered in \cite{hradil2004}:
\begin{equation}
\label{multilik}
{\cal L}(\rho | n) = \frac{N!}{\prod_{i=1}^m n_i} \prod_{j=1}^m p_j(\rho)^{n_j} = \frac{N!}{\prod_{i=1}^m n_i} \prod_{j=1}^m \trace{E_j \rho}^{n_j},
\end{equation}
where $n_j$ is the number of occurrences of the outcome $j$ and $N=n_1 + \dots + n_m$. \\
Another model, used in photonic tomography \cite{altepeter2005}, that takes into account the noise detection proposed by \cite{james2001} is
\begin{equation}
\label{gausslik0}
{\cal L}(\rho | n) = \frac{1}{N_{norm}} \prod_{j \in {\cal I}} \exp \left( -\frac{1}{2} \frac{(N\trace{E_j \rho} - n_j)^2}{N\trace{E_j \rho}} \right),
 \end{equation}
where $N_{\mbox{norm}}$ is a normalization constant. 
 
Solving \refe{mle} for a general non-linear ${\cal L} (\rho)$ is a challenge on its own. Numerical methods just  for specific likelihoods have been developed. In case of \refe{multilik}, the $R \rho R$ algorithm and variants  \cite{hradil2004,rehacek2007} have shown good practical performance. For the likelihood \refe{gausslik0} a common method is to reparameterize the density matrix:
\begin{equation}
\rho = \frac{T^{\dagger}T}{\trace{T^{\dagger}T}},
\end{equation}  
where $T$ is an upper triangular matrix, which results  in an unconstrained optimization problem
 in the entries of $T$.
 Although unconstrained, this new problem in $T$ is non-convex and plagued with many local maxima. 
Only recently \cite{goncalves2012} it has been proved that these local maxima are all global. 

Regardless the method used to solve \refe{mle}, whenever we do not have a tomographically 
complete set of measurements or some data is missing (incomplete data), 
there is a convex set of maximizers of ${\cal L}(\rho | n)$, that is, a convex set of matrices for which the likelihood achieves its maximum value. Then the intersection of this set with the set of density matrices may be a non-empty convex set and, in this case, we have more than one solution for \refe{mle}. Each of these solutions is compatible with the data, in the sense that minimizes some relative distance between the observed data $f_j$ and the probabilities predicted by quantum mechanics $\trace{E_j \rho}$. However, these solutions differ on how to fit the unmeasured probabilities. 

One way to choose a unique solution from the set of ML (Maximum Likelihood) solutions,
 in this case, is to apply the MaxEnt method constrained to the set of ML solutions. To accomplish this task, after we have one ML solution, say $\rho_{ML}$, we know that
\begin{equation}
\trace{E_i \rho} = \bar{p}_i \equiv \trace{E_i \rho_{ML}}, \ \ \forall i \in {\cal I}, \ \ \forall \rho \in {\cal S}_{ML},
\end{equation}
where ${\cal S}_{ML}$ is the ML solution set. Therefore,  we can obtain the Maximum Entropy solution among the ML solutions solving the problem \refe{maxent} switching $f_i$ by $\bar{p}_i$ for all $i \in {\cal I}$. This is the Maximum Likelihood - Maximum Entropy (MaxLik-MaxEnt) \cite{rehacek2005} estimate for $\rho$. Other possible approach is the joint maximization of likelihood and entropy through a Lagrangian function \cite{hradil2012}.

\section{Numerical simulations: VQT$_{\infty}$ and MaxLik-MaxEnt}\label{simsec}

\begin{figure}[!h]
\epsfig{file=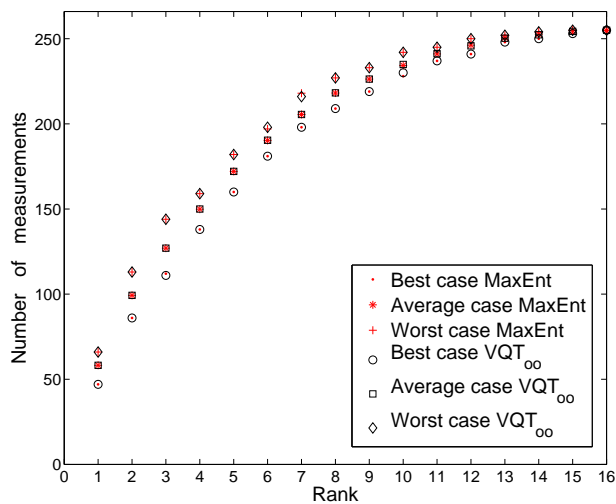, width=9.3cm}
\caption{(Color online) Convergence of VQT$_{\infty}$ and MaxEnt increasing the number of measurements.}
\label{fig1}
\end{figure}

In order to compare the VQT$_{\infty}$ and the MaxEnt approach, for each rank,
 we sampled 100 uniformly 
distributed density matrices according to the Haar measure \cite{karol2011}, representing four-qubits states.
 We fixed a SIC-POVM base \cite{renes2004} and, for each rank, Fig. \ref{fig1} shows the worst,
 the average and the best number of measurements needed for each method to converge to the reference state.
 We considered that a method converges if the trace distance to the real state is less than $10^{-4}$.
 Firstly we consider noise-free data.
As one can see, the required number of measurements to converge is almost the same for MaxEnt and VQT$_{\infty}$ methods. Fig. \ref{fig1} also shows that the worst case for each method does not exceeds the ${\cal O}(r d \log d)$ number of measurements ($r$ is the rank) mentioned in compressed sensing works \cite{gross2010} in quantum tomography. 

\begin{figure}[!h]
\centering
\epsfig{file=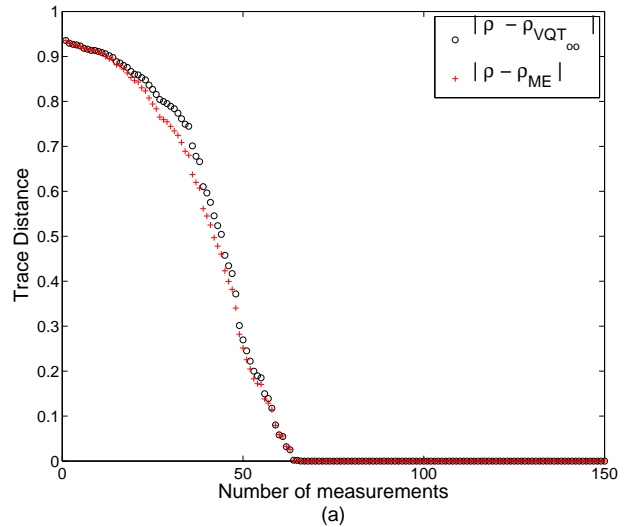, width=9.3cm}
\epsfig{file=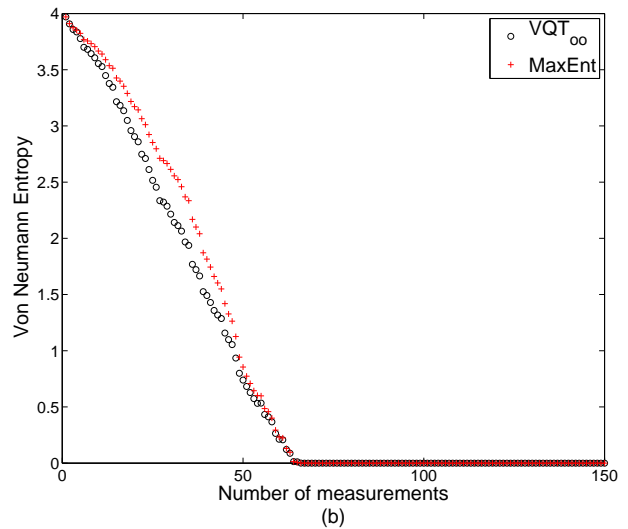, width=9.3cm}
\caption{(Color online)(a): Trace distance to the real rank-1 state for VQT$_{\infty}$ and MaxEnt. (b): von Neumann entropy of VQT$_{\infty}$ and MaxEnt estimates.}
\label{fig2}
\end{figure}

\begin{figure}[!h]
\centering
\epsfig{file=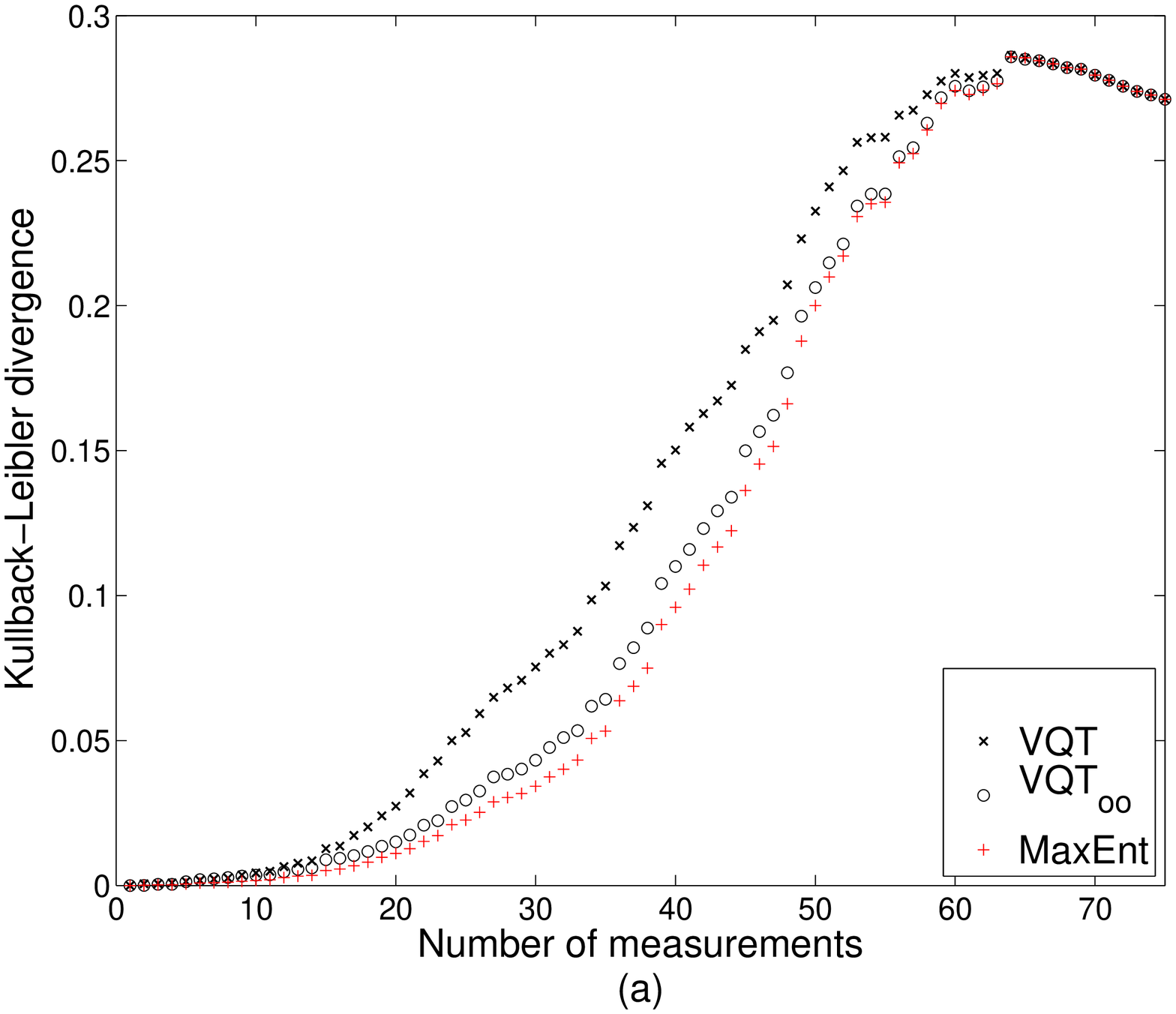, width=9.3cm}
\epsfig{file=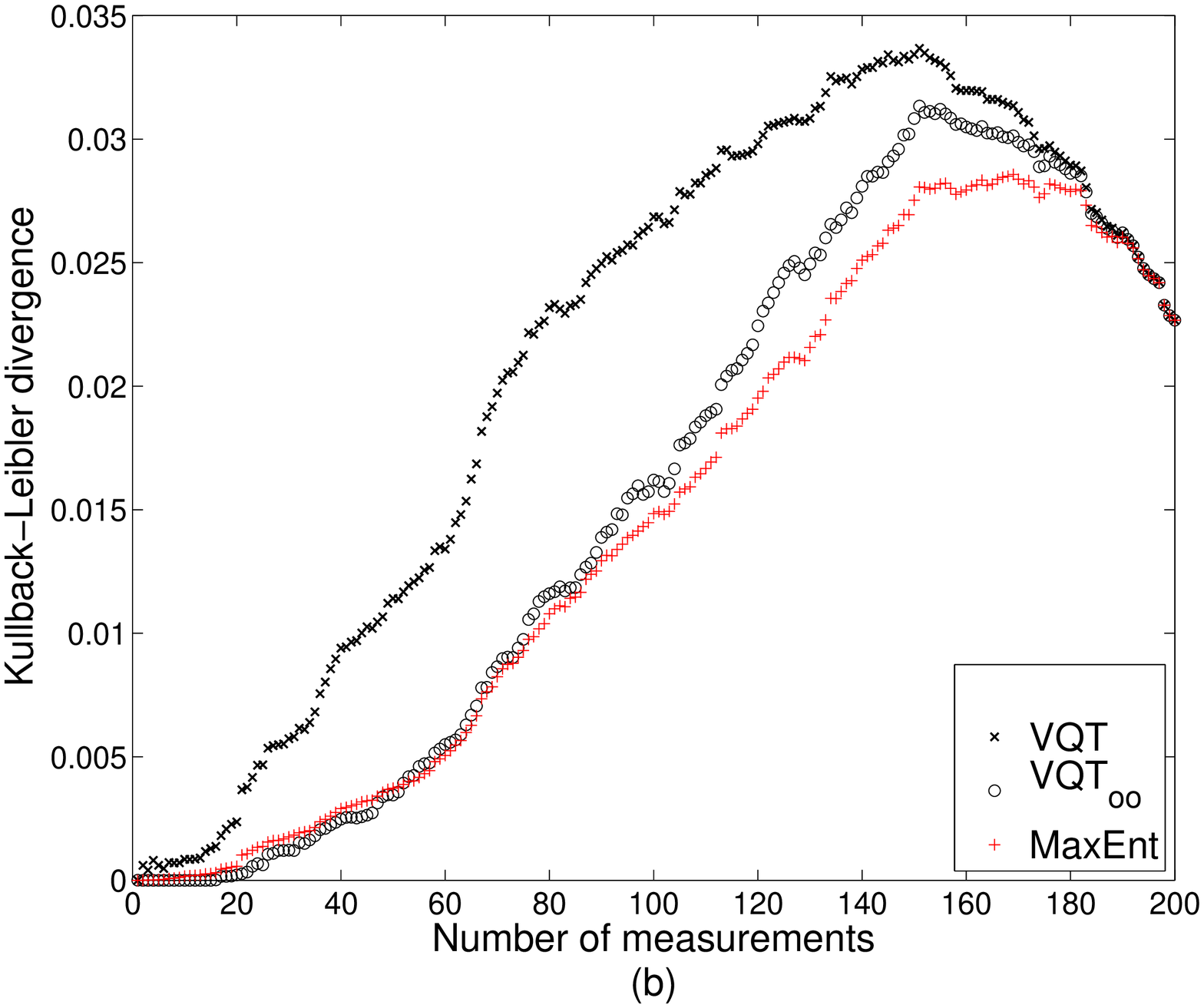, width=9.3cm}
\caption{(Color online) Average Kullback-Leibler divergence of $\tilde{c}$ to the uniform distribution. (a) rank one, (b) rank six.}
\label{klunif}
\end{figure}

\begin{figure}[!h]
\centering
\epsfig{file=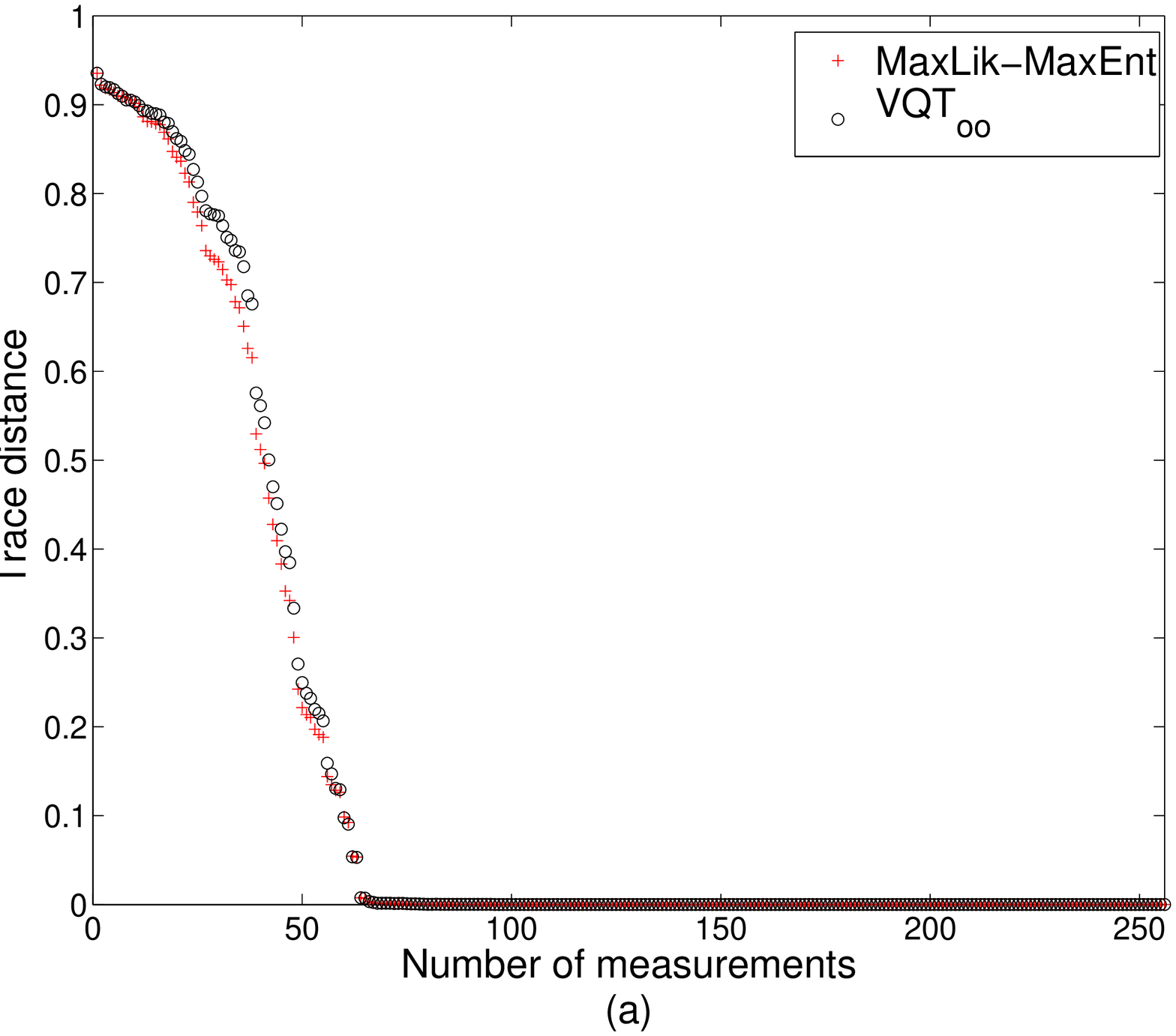, width=9.3cm}
\epsfig{file=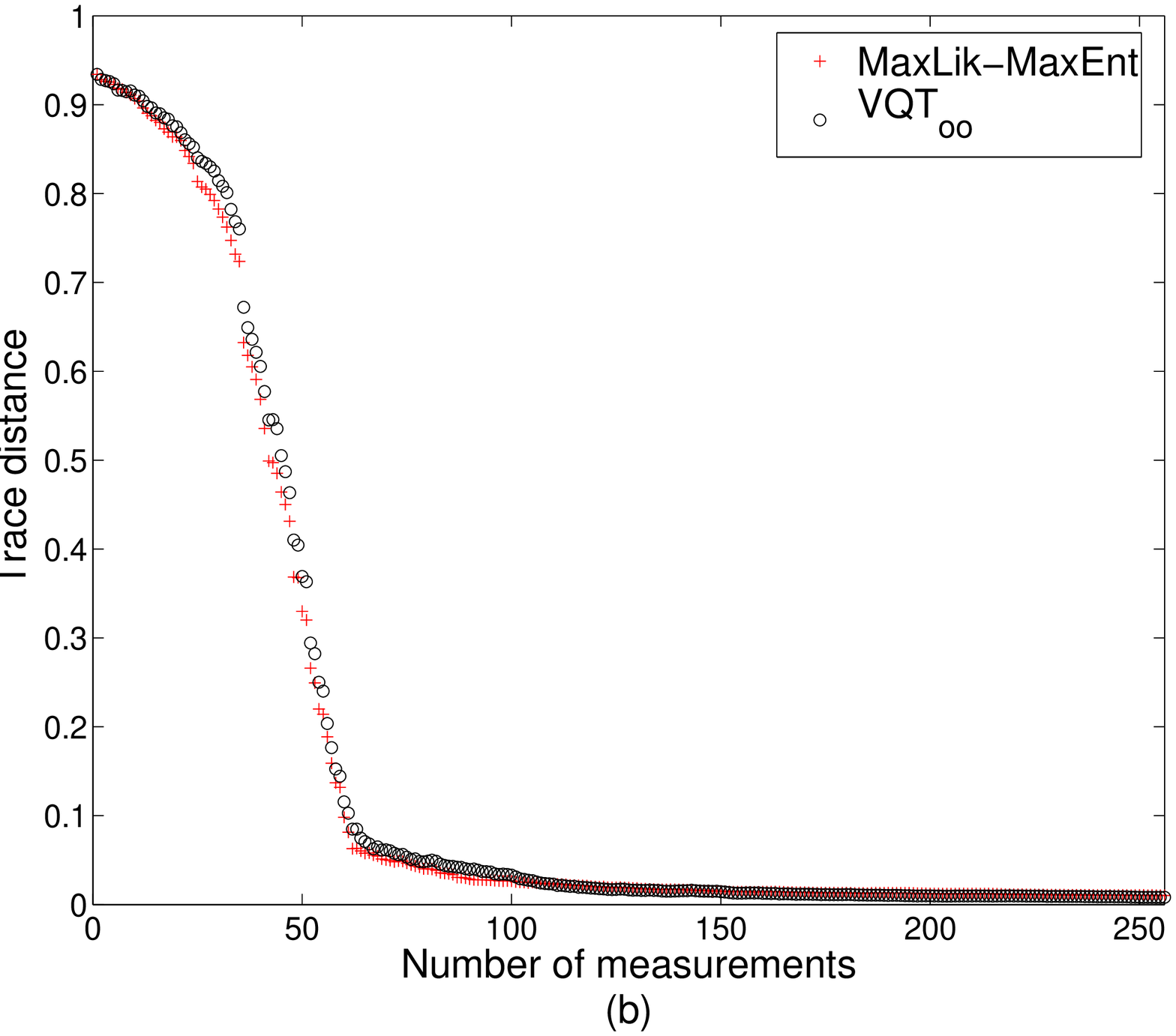, width=9.3cm}
\caption{(Color online) (a): Trace distance to the reference rank-1 state with probabilities perturbed by a Gaussian noise. (b): Trace distance to the reference state with 5\% uniform noise.}
\label{dist_mlme_vqt}
\end{figure}

Now, we compare the convergence of both methods in terms of average trace distance and average entropy for rank one states as the measurements increase, illustrated in Fig. \ref{fig2}. Again, one can see a similar behavior between MaxEnt and VQT$_{\infty}$, the former with a slightly smaller distance and greater entropy than the second.\\
The impact of the proposed modification \refe{vqtinf0} comes up when we compare the distance of $\tilde{c}$, the vector of remaining probabilities, to the uniform vector whose entries are defined in \refe{uvec}.
 Fig.  \ref{klunif}  shows, for ranks one and six, the average Kullback-Leibler divergence to the uniform distribution. As one can see, the VQT$_\infty$ is closer to the MaxEnt than the original VQT.
 When the number of measurements is sufficient to determine uniquely the state, 
 then we observe the agreement of the methods.
To assess the convergence properties in a more realistic scenario we introduced two kinds of error in the true probabilities, one using a Gaussian perturbation with zero mean and standard deviation $10^{-6}$ and the other considering values uniformly distributed in an interval of 5\% deviation of the true probabilities. Since the equations \refe{linear_eq} may not be satisfied for noisy data,
 we use the MaxLik-MaxEnt method in comparison with VQT$_{\infty}$. For the likelihood function we use a variant of \refe{gausslik0}. In Fig. \ref{dist_mlme_vqt}, we plot the average trace distance depending on the number of measurements for random rank one states for the two kinds of error.
These numerical simulations corroborate the relationship between VQT$_{\infty}$ and MaxEnt methods already proved in Section \ref{vqteigb} for eigenbasis measurements. The new proposed formulation \refe{vqtinf}, as the MaxEnt approach, tries to fit the unmeasured probabilities  
the most uniformly as possible. We see that the results for these methods are quite close and the VQT$_{\infty}$ has the advantage of the linear SDP programming against non-linear optimization problems of MaxEnt. 
Each tomography using VQT takes no more than five seconds whereas MaxLik-MaxEnt tomography spends about twelve seconds in the worst case. The simulations were done in a Intel Core 2 Duo, 2 GB RAM computer, in MATLAB, using Yalmip/SEDUMI \cite{yalmip,sedumi} to model and solve the SDP problems.

\section{Final remarks}\label{finalsec}
We proposed a variant of the VQT method for quantum tomographies with incomplete information, namely VQT$_{\infty}$,  that tries to fit the unmeasured probabilities the most uniformly as possible.
 As a consequence, we showed that the VQT$_{\infty}$ has a quite similar behavior to 
the well-known MaxEnt approach in the noise-free case, and to the MaxLik-MaxEnt in the presence of
 noise. This claim was confirmed by the numerical simulations and theoretically for 
the case of eigenbasis measurements. 

Thus, using the VQT$_{\infty}$ method one obtains an estimate for the density matrix as unbiased as the MaxEnt and with the advantage of dealing with linear SDP problems that can be solved efficiently by methods that have polynomial complexity, a useful property in the tomography of increasing large quantum systems.

\section*{ACKNOWLEDGMENTS}
The authors acknowledge the anonymous referees for suggestions to improve this paper.

This work was partially supported by the Brazilian research agencies FAPESP, FAPEMIG, CNPq and INCT-IQ (National Institute of Science and Technology for Quantum Information).

\bibliographystyle{apsrev4-1}
\bibliography{vqt_maxent}

\end{document}